\documentclass[%
 aip,
 amsmath,amssymb,
 reprint,%
]{revtex4-1}

\usepackage{graphicx}
\usepackage{dcolumn}
\usepackage{bm}

\usepackage[utf8]{inputenc}
\usepackage[T1]{fontenc}
\usepackage{mathptmx}
\usepackage{etoolbox}

\makeatletter
\def\@email#1#2{%
 \endgroup
 \patchcmd{\titleblock@produce}
  {\frontmatter@RRAPformat}
  {\frontmatter@RRAPformat{\produce@RRAP{*#1\href{mailto:#2}{#2}}}\frontmatter@RRAPformat}
  {}{}
}%
\makeatother
\begin{document}

\preprint{AIP/123-QED}

\title[Chiral standing spin waves in skyrmion lattice]{Chiral standing spin waves in skyrmion lattice}

\author{Andrii~S.~Savchenko}
\email{a.savchenko@fz-juelich.de} 
 \affiliation{Peter Gr\"unberg Institut and  Institute for Advanced Simulation,  Forschungszentrum J\"ulich and JARA, D-52425  J\"ulich, Germany}
 \affiliation{Donetsk Institute for Physics and Engineering, National Academy of Sciences of Ukraine, 03028 Kyiv, Ukraine}

\author{Vladyslav~M.~Kuchkin}
 \affiliation{Peter Gr\"unberg Institut and  Institute for Advanced Simulation,  Forschungszentrum J\"ulich and JARA, D-52425  J\"ulich, Germany}

\author{Filipp~N.~Rybakov}
\affiliation{Uppsala University, SE-75120 Uppsala, Sweden}
\affiliation{KTH Royal Institute of Technology, SE-10691 Stockholm, Sweden}

\author{Stefan~Bl\"ugel} 
\affiliation{Peter Gr\"unberg Institut and  Institute for Advanced Simulation,  Forschungszentrum J\"ulich and JARA, D-52425  J\"ulich, Germany}

\author{Nikolai~S.~Kiselev}
\affiliation{Peter Gr\"unberg Institut and  Institute for Advanced Simulation, Forschungszentrum J\"ulich and JARA, D-52425  J\"ulich, Germany}

\date{\today}

\begin{abstract}
This work studies the resonance excitations of the three-dimensional skyrmions lattice in the finite thickness plate of an isotropic chiral magnet using spin dynamics simulations.
We found that the absorption spectra and resonance modes differ from those predicted by the two-dimensional model and the model of the unconfined bulk crystal.
The features observed on the spectra can be explained by the formation of chiral standing spin waves, which, contrary to conventional standing spin waves, are characterized by the helical profile of dynamic magnetization of fixed chirality defined by the Dzyaloshinskii-Moriya interaction.
In this case, the dynamic susceptibility becomes a function of the plate thickness, which gives rise to an interesting effect that manifests itself in periodical fading of the intensity of corresponding modes and makes excitation of these modes impossible at specific thicknesses.
\end{abstract}

\maketitle

\section{\label{sec:level1}Introduction}

The shape and size of magnetic samples defines many of their static and dynamic properties.
The most representative examples is so-called \textit{bubble materials}~\cite{Bobeck_75, Malozemoff_79} -- thin ferromagnetic films with strong perpendicular anisotropy.
The size of the magnetic domains in these systems depends on the film thickness.
The variation of the thickness changes the energy balance between demagnetizing fields effects, Heisenberg exchange, and magnetic anisotropy.
The magnetic phases observed in the bulk crystals and thin films of bubble materials are very different.
The ground state of very thin films with thickness comparable to the domain wall width is the homogeneous ferromagnetic state, while for thicker films, the multidomain states -- stripes or bubbles become more energetically favorable.

Another example of the systems where shape of the samples plays a crucial role is so-called chiral magnets where the main properties are defined by the competition between the Heisenberg exchange interaction and chiral Dzyaloshinskii-Moriya interaction (DMI)~\cite{Dzyaloshinskii,Moriya}. 
At zero external magnetic field, in bulk crystals of B20-type FeGe~\cite{Yu_11, Kovacs_17, Du_18, Yu_18}, MnSi~\cite{Ishikawa_84,Grigoriev_06,Mulbauer_09,Yu_15}, Fe$_{1-x}$Co$_x$Si~\cite{Yu_10, Park_14} and others compound from that class~\cite{Shibata_13}, the ground state of the system is the helical spiral. 
With increasing external magnetic fields, the spiral continuously transforms into the cone phase and than into the field saturated state.
These three phases are the main phases observed in the bulk crystals of isotropic chiral magnets in a whole range of temperatures. 
There are, however, a few exceptions, for instance, a tiny pocket near the critical temperature on the phase diagram of the most of the bulk chiral magnets known as the anomalous phase~\cite{Ishikawa_84, Grigoriev_06, Mulbauer_09} and additional phases which are observed at low temperature in Cu$_2$OSeO$_3$~\cite{Pfleiderer_18, Pappas_18}.
The experimental observations in thin films of chiral magnets show the emergence of another phase -- the skyrmion lattice, which remains stable in a wide range of magnetic fields~\cite{Yu_11, Kovacs_17, Du_18, Yu_18,Yu_15,Yu_10, Park_14,Shibata_13}. 
The theoretical explanation of this phenomenon is based on the effect known as the chiral surface twist~\cite{Meynell_14,Du_15}, which leads to a significant skyrmion energy gain in the case of the films with the thickness of a few periods of helical modulations~\cite{Rybakov_13, Rybakov_16}. 
The energy gain provided by the chiral twist near the surface reduces with the film thickness and above critical thickness the skyrmion phase becomes energetically unfavorable~\cite{Rybakov_16}.

It is natural to expect that besides the static properties, the system's dimensionality and the edge effects also significantly affect the dynamic properties of the chiral magnets.
For instance, it is known that in pure ferromagnetic films at particular boundary conditions~\cite{Melkov_book} the spin-wave resonance~\cite{Kittel_58, Seavey_58} can take place. 
In chiral magnets, the presence of DMI causes an effective ``pinning'' of the spins at the boundaries~\cite{Rohart_13, Meynell_14, Puszkarski_75, Garst_17, Kovalev_18}. 
The latter can lead to the formation of the standing spin-waves (SSW), where all the spins precess with identical frequency, but the relative phase and the amplitude of the spin precessions vary across the thickness. At the same time, the positions of the minima and maxima of the amplitude -- nodes and antinodes, respectively, remain fixed in time.
Analytical models based on the linear approximation suggest that SSW can arise in chiral magnets in the field saturated state and in the helical spin spiral state ~\cite{Garst_17, Stasinopoulos_17, Maisuradze_17, Aqeel_21}.
On the other hand, in some works it was argued that SSW with fixed in time positions of the antinodes do not exist in chiral magnets~\cite{Zingsem_19} or at least cannot be considered as conventional SSW~\cite{Che_21}.
One of the aims of this work is to confirm the presence of SSW in chiral magnets and highlight some essential aspects that have not been discussed in the previous studies.

Here we present the results of the theoretical study of the spin-wave resonance in the three-dimensional (3D) skyrmions lattice (SkL) representing statically stable configuration in the plate of isotropic chiral magnets. We show that the spectra of resonance frequencies, in this case, differ from the 2D model and the additional resonance modes can be explained in terms of chiral SSW.

\section{Model}

We consider the classical spin model of the simple cubic lattice described by the following Hamiltonian:
\begin{align}
E\!=\!- J \sum_{\left\langle ij\right\rangle }
 \mathbf{n}_i  \cdot  \mathbf{n}_j - 
\sum_{\left\langle ij\right\rangle } 
\!\mathbf{D}_{ij}  \cdot  [\mathbf{n}_i\! \times\!  \mathbf{n}_j]   - \! \mu_\mathrm{s} \mathbf{B}\!\sum_{i}\mathbf{n}_i,         
\label{E_tot}
\end{align}
where $\mathbf{n}_i=\mathbf{M}_i/\mu_\textrm{s}$ is a unit vector of the magnetic moment at site $i$,
$\mu_\mathrm{s}$ is the value of magnetic moment of the lattice site,
$\left\langle ij\right\rangle$ means summation over all nearest-neighbour pairs,
$J$ is the exchange coupling constant and $\mathbf{D}_{ij}$ is the Dzyaloshinskii-Moriya vector defined as $\mathbf{D}_{ij}=D \, \mathbf{r}_{ij}$, 
$D$ is the DMI scalar constant and $\mathbf{r}_{ij}$ is the unit vector between sites $i$ and $j$.
The external magnetic field, $\mathbf{B}(t)=\mathbf{B}_\mathrm{DC}+\mathbf{B}_\mathrm{AC} (t)$,
contains the static (DC) and time-dependent (AC) components.
To ensure the generality of the results, the external magnetic fields and distances are given in reduces units with respect to the saturation field $B_\mathrm{D}=D^2/(\mu_\mathrm{s}J)$ and equilibrium period of helical modulations in the continuum limit,  $L_\mathrm{D}=2\pi aJ/D$, respectively.
We excluded the dipole-dipole interactions since they do  not affect the discussed phenomena significantly.

\begin{figure}
\centering
\includegraphics[width=8cm]{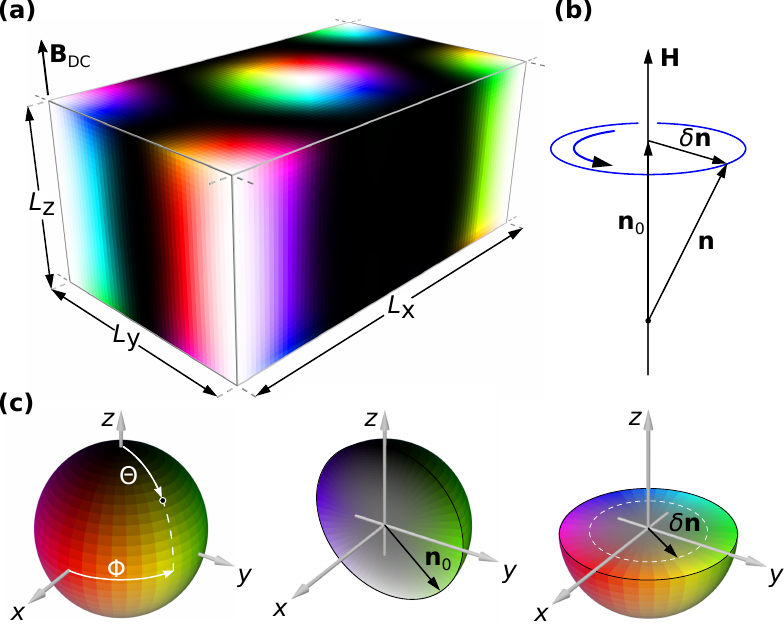}
\caption{(a) The simulated domain with the size fitting the equilibrium period of 3D skyrmion lattice with periodic boundary condition in $xy$ plane. The magnetization field $\mathbf{n}_0 (x,y,z)$ is represented by the color code. The static magnetic filed is perpendicular to the film plane, $\mathbf{B}_\mathrm{DC}\parallel \mathbf{e}_\mathrm{z}$. (b) In the excited state, the spins oscillate about the direction of effective field and at resonance can be decomposed into static component $\mathbf{n}_0$ and dynamic component $\delta \mathbf{n}$. (c) Color code for used for unit vector field of magnetization $\mathbf{n}$ and dynamic component of magnetization $\delta \mathbf{n}$. 
}
\label{Fig1}
\end{figure}

We consider extended film of finite thickens with periodic boundary conditions (PBC) in the $xy-$plane and free boundaries along the $z$-axis [Fig.~\ref{Fig1}(a)].
The shape and size of the simulated domain in the $xy-$plane is chosen to fit the equilibrium size of the unit cell of the hexagonal skyrmion lattice, $L_\mathrm{x}/L_\mathrm{y}=78/45\approx \sqrt{3}$.
The thickness of the film $L_\mathrm{z}$ is varying by changing the number of atomic layers, $d$, from 1 to 140, which for chosen coupling parameters corresponds to $L_\mathrm{z}=0$ and $L_\mathrm{z}=4 L_\mathrm{D}$.
The static magnetic field is always perpendicular to the film plane, $\mathbf{B}_\mathrm{DC}\parallel \mathbf{e}_\mathrm{z}$.
In the following we study both cases of in-plane and out-of-plane alternate magnetic fields:  $\mathbf{B}_\mathrm{AC}\perp \mathbf{B}_\mathrm{DC}$ and $\mathbf{B}_\mathrm{AC}\parallel \mathbf{B}_\mathrm{DC}$.

The dynamics of magnetization is described by the Landau-Lifshitz-Gilbert (LLG) equation, which can be written in the dimensionless form:
\begin{equation}
\frac{\partial \mathbf{n}_i}{\partial t} = -\frac{1}{1+\alpha^2} (\mathbf{n}_i \times \mathbf{H}_i + \alpha  \mathbf{n}_i \times (\mathbf{n}_i \times \mathbf{H}_i)),
\label{LLG1}
\end{equation}
where $t$ is a dimensionless time scaled by $\gamma E_0/\mu_\mathrm{s}$, where $\gamma$ is the gyromagnetic ratio and $E_0$ is the reference energy, which we set equal to exchange constant $J$, since the  Heisenberg exchange is the leading energy term in ~\eqref{E_tot},
$\alpha$ is the Gilbert damping, 
and $\mathbf{H}_i=-E_0^{-1}{\partial E}/{\partial \mathbf{n}_i}$ is a dimensionless effective field on the $i$th lattice site.
The solution of LLG equation provides time dependent magnetization vector field $\mathbf{n}(t)$ that in case of small oscillations can be always decomposed into static, $\mathbf{n}_0$, and dynamic, $\delta \mathbf{n}(t)$, components: $\mathbf{n}(t)=\mathbf{n}_0+\delta \mathbf{n}(t)$ [see Fig.~\ref{Fig1}(b)].
For visualisation of the fields $\mathbf{n}_0$ and $\delta \mathbf{n}$, we use the color scheme explained by Fig.~\ref{Fig1}(c).

\begin{figure*}
\centering
\includegraphics[width=17cm]{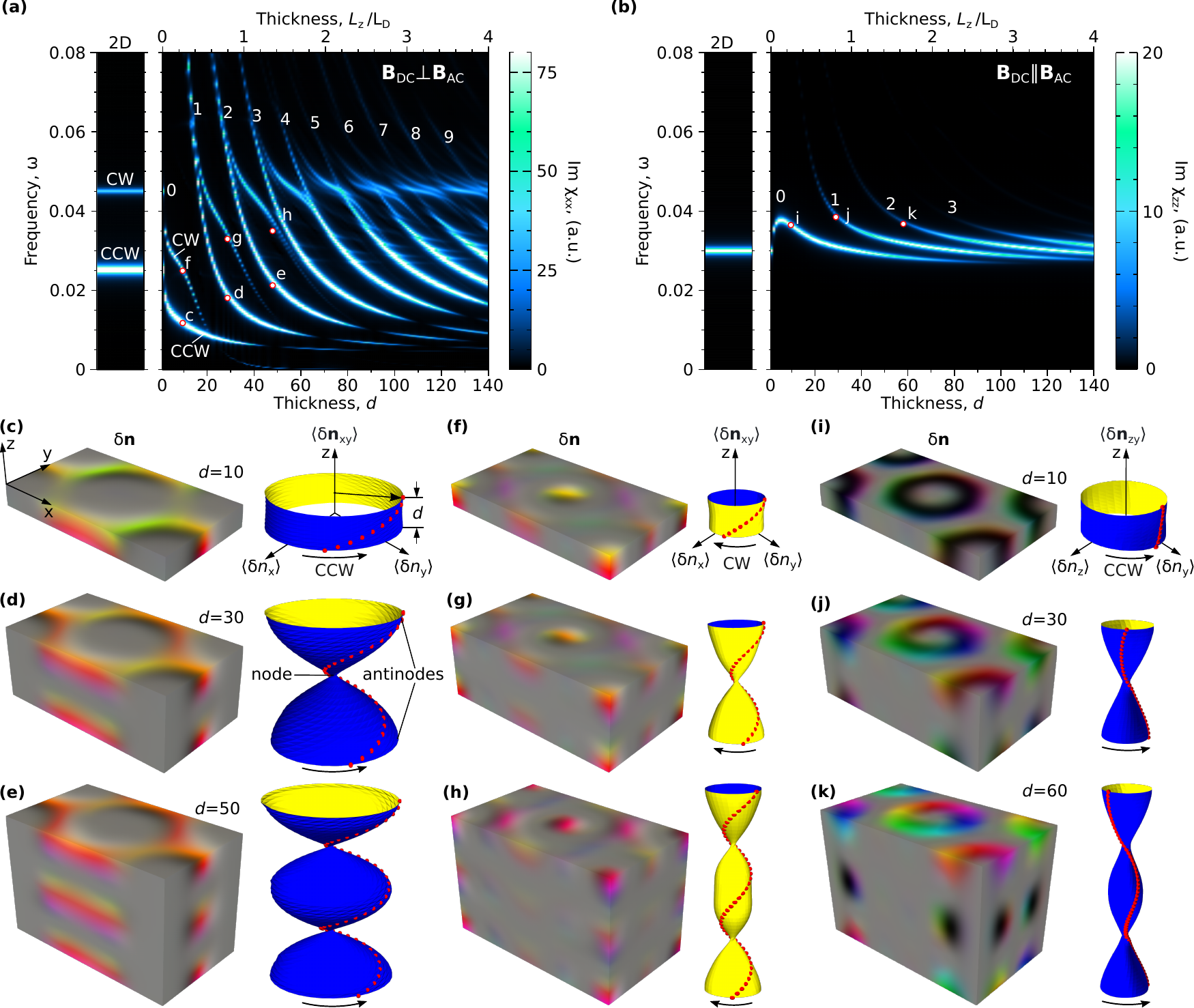}
\caption{
(a) Absorption spectra $\mathrm{Im}\chi_{xx}(\omega)$ of 2D (left) and 3D skyrmion lattice (right) for different thicknesses $d=L_z/a$ of chiral magnet film under in-plane exciting field $\mathbf{B}_\mathrm{AC}\parallel \mathbf{e}_\mathrm{x}$; 
white digits correspond to indexes of modes; red-white dots indicate values for visualization of dynamic magnetization distributions $\delta \mathbf n(x,y,z)$ and spin-wave profiles depicted below. 
(b) The thickness dependent spectra of  $\mathrm{Im}\chi_{zz}(\omega)$ in case out-of-plain exciting magnetic field $\mathbf{B}_\mathrm{AC}\parallel \mathbf{e}_\mathrm{z}$. 
(c)-(e) Snapshots of $\delta \mathbf n(x,y,z)$ in simulated domain at fixed time and profiles of average value $\langle\delta \mathbf {n}_{xy}\rangle$ along film thickness for low-frequency modes with indexes $p=0,1,2$ correspondingly for $\mathbf{B}_\mathrm{AC}\parallel \mathbf{e}_\mathrm{x}$. 
(f)-(h) Snapshots of $\delta \mathbf n(x,y,z)$ and  $\langle\delta \mathbf {n}_{xy}\rangle$ for high-frequency modes with indexes $p=0,1,2$ for $\mathbf{B}_\mathrm{AC}\parallel \mathbf{e}_\mathrm{x}$; white dashed line indicates average region for $\langle\delta \mathbf {n}_{xy}\rangle$;
(i)-(k) Snapshots of $\delta \mathbf n(x,y,z)$ and $\langle\delta \mathbf {n}_{zy}\rangle$ of modes with indexes $p=0,1,2$ for $\mathbf{B}_\mathrm{AC}\parallel \mathbf{e}_\mathrm{z}$.
}
\label{SKL}
\end{figure*}

The calculation of the absorption spectra in Figs.~\ref{SKL}(a) and ~\ref{SKL}(b) is based on Fourier analysis of the dynamic component of magnetization induced by an external magnetic field pulse (see Appendix \ref{Appendix_A} for details).
The absorption spectra in Figs.~\ref{SKL}(a) and ~\ref{SKL}(b) can be thought of as the system response to the external magnetic field oscillating with the frequency $\omega$.
The picks of the imaginary component of complex dynamic susceptibility, $\mathrm{Im}\chi (\omega)$,  correspond to those eigenfrequencies of the system, which can be excited in skyrmion lattice by the uniform AC field.

\section{Results}

\subsection{\label{sec:level2}The resonant spectra}

First, we consider the case of $\mathbf{B}_\mathrm{AC}\perp \mathbf{B}_\mathrm{DC}$.
The dynamic susceptibility, $\mathrm{Im}\chi_{xx}(\omega)$, as a function of the film thickness is shown in Fig.~\ref{SKL}(a).
In contrast to the 2D case~\cite{Mochizuki_2012} (see left panel), the spectrum for the 3D SkL has larger number of resonance modes marked by index $p$.
Each mode has high- and low-frequency branches which merge with decreasing thickness. %
The exception is the pair of modes with $p=0$, which in the limit of the very thin film converge to clockwise (CW) and counter-clockwise (CCW) precession modes of 2D SkL \cite{Mochizuki_2012}.
Another remarkable feature of the pair of modes $p=0$ is the presence of the crossover point at the thickness of about $L_\mathrm{z}=0.62L_\mathrm{D}$ ($d=21$), above which the frequency of the CW mode becomes lower than that for the CCW mode.
This effect is absent for the modes with $p\geq1$.

Modes belonging to the high-frequency branches have lower intensity which quickly decays with the thickness.
For $p>2$, the high-frequency and low-frequency branches overlap and hybridize, which makes identifying a particular mode challenging.
Because of that, in the following, we focus on the modes with $p\leq2$ and film thicknesses below $2L_\mathrm{D}$ only.

The functional dependence of the resonance frequency on the thickness is $\omega_{p}(d) \sim p/d$, which is most prominent for low-frequency modes. 
Such behavior of the resonance frequencies resembles the behavior of SSW.
Below we show that observed phenomena indeed can be explained in terms of SSW. 
Furthermore, we demonstrate that the presence of the DMI plays an essential role in this phenomenon and gives rise to several interesting effects that essentially distinguish SSW in chiral magnets from SSW observed in non-chiral magnets.

The dependence of $\mathrm{Im}\chi_{zz}(\omega)$ on the film  thickness for $\mathbf{B}_\mathrm{AC}\parallel \mathbf{B}_\mathrm{DC}$ is provided in Fig.~\ref{SKL}(b).
In the limiting case of one layer, $d=1$ (see left panel for 2D SkL), there is only one resonance mode -- so-called breathing mode~\cite{Mochizuki_2012}.
With increasing thickness the number of resonance modes increases [Fig.~\ref{SKL}(b)].
Similar to the case of $\mathbf{B}_\mathrm{AC}\perp \mathbf{B}_\mathrm{DC}$ the resonance frequency and the intensities of that modes decrease with increasing thickness. On the other hand, the functional dependencies for $\omega_p (d)$ for the cases of in-plane and out-of-plane AC fields are different.


\subsection{Visualization of resonance modes}

To illuminate the physical origin of the spectra, shown in Figs.~\ref{SKL}(a) and ~\ref{SKL}(b), we analyze the profile of the dynamic component of magnetization $\delta \mathbf{n}$ for different resonance modes.
For both case of in-plane and out-of-plane AC field we have chosen three representative thicknesses and the frequencies corresponding to the first three modes, $p=0$, 1, and 2, see red hollow circles in Figs.~\ref{SKL}(a) and ~\ref{SKL}(b).
%
On the left panel in Figs.~\ref{SKL}(c)-(k), we show representative snapshots of $\delta \mathbf{n}(x,y,z)$, for each of that modes.
Since $\delta \mathbf{n}$ is not a unit vector field, we use a special color scheme reflecting both the direction and the length of the vectors [see Fig.~\ref{Fig1}(c)].
When $|\delta \mathbf n|\rightarrow0$ the color converges to a gray and for $|\delta \mathbf n|\rightarrow \delta  n_\mathrm{max}$ the saturation of the color reaches maximum value.
Here $\mathbf \delta n_\mathrm{max}$ is the maximal value of the $|\delta \mathbf n|$ over the whole domain.
The details of $\delta \mathbf{n}(x,y,z)$ calculations are provided in Appendix \ref{Appendix_A}.

%
In the case of $\mathbf{B}_\mathrm{AC}\perp \mathbf{B}_\mathrm{DC}$, for low-frequency modes [see Figs.~\ref{SKL}(c)-(e)], the  $|\delta \mathbf{n}|$  is maximal in the inter-skyrmion area  where magnetization is parallel to the $\mathbf{B}_\mathrm{DC}$.
In case of high-frequency modes [Figs.~\ref{SKL}(f)-(h)], the maximum of the $|\delta \mathbf{n}|$ is mainly located in the core of the skyrmions, where magnetization is antiparallel to the $\mathbf{B}_\mathrm{DC}$. 
Since the high-frequency modes are localized in a much smaller volume their intensities in the absorption spectrum are much lower than that of low-frequency modes.

For $\mathbf{B}_\mathrm{AC}\parallel \mathbf{B}_\mathrm{DC}$, $|\delta \mathbf{n}|$ takes its maximal values in the skyrmion shell where magnetization is mainly lying in the plane, $\mathbf{n}\perp\mathbf{e}_\mathrm{z}$.
Nevertheless, the distribution of $\delta \mathbf{n}$ across the thickness is similar to the case of $\mathbf{B}_\mathrm{AC}\perp \mathbf{B}_\mathrm{DC}$ and characterized by the presence of clearly visible minima and maxima. 
In the right panel of Figs.~\ref{SKL}(c)-(k) we provide the profiles of the dynamic magnetization averaged in certain area $\Omega$ in the $xy$-plane, $\langle\delta \mathbf{n}\rangle\equiv\langle\delta \mathbf{n}\rangle(t,z)=\sum_{(x,y)\in\Omega}\delta \mathbf{n}(t,\mathbf{r})$, which aim making the presence of the minima and maxima more evident. The details of the $\langle\delta \mathbf{n}\rangle$ profiles calculation and choosing the area $\Omega$ are provided in Appendix~\ref{Appendix_A}.
The chains of red dots correspond to the snapshots of the $\langle\delta \mathbf{n}\rangle (z)$ at fixed moment in time.
The yellow-blue surfaces of revolution are created by a set of such snapshots taken with equal time intervals (see Appendix \ref{Appendix_A}). 

The dynamic magnetization surfaces have nodes and antinodes. The number of nodes equals to the index of the corresponding mode, $p$. 
A major difference between the low-frequency and the high-frequency modes is the opposite sense of $\langle\delta \mathbf {n}_{xy}\rangle$ rotation.
The vectors $\langle\delta \mathbf {n}_{xy}\rangle$ rotate counterclockwise (CCW) in the case of low-frequency modes (see Supplementary Movies 1-3) and clockwise (CW) in the case of high-frequency modes (see Supplementary Movies 4-6).
For breathing modes the spins rotate CCW about chosen precession axis (see Supplementary Movies 7-9).
The sense of $\langle\delta \mathbf {n}\rangle$ rotation is defined by the LLG equation and orientation of the projection plane normal with respect to effective field $\mathbf{H}$.

Since the position of the nodes and antinodes remain stable in time, for  the case of $\mathbf{B}_\mathrm{AC}\perp \mathbf{B}_\mathrm{DC}$, we identify  the corresponding excitation modes as \textit{chiral} SSW.
We use term ``chiral'' because the profile of the dynamic magnetization $\langle\delta \mathbf {n}\rangle(z)$ has a helical profile with fixed chirality.
On the contrary, in case of SSW in ferromagnet without DMI the vectors $\langle\delta \mathbf {n}\rangle$ always lie in the same plane~\cite{Melkov_book, Stancil_book}.
Moreover, for the case of $\mathbf{B}_\mathrm{AC}\perp \mathbf{B}_\mathrm{DC}$ in both cases of low-frequency and the high-frequency modes depicted in Figs.~\ref{SKL}(c)-(h) the chirality of the $\langle\delta \mathbf {n}\rangle(z)$ profile across the thickness is defined by the chirality of the DMI.
The period of such chiral modulation of the $\langle\delta \mathbf {n}\rangle(z)$ profile equals $L_\mathrm{D}$.

\begin{figure*}[ht!]
\centering
\includegraphics[width=17.8cm]{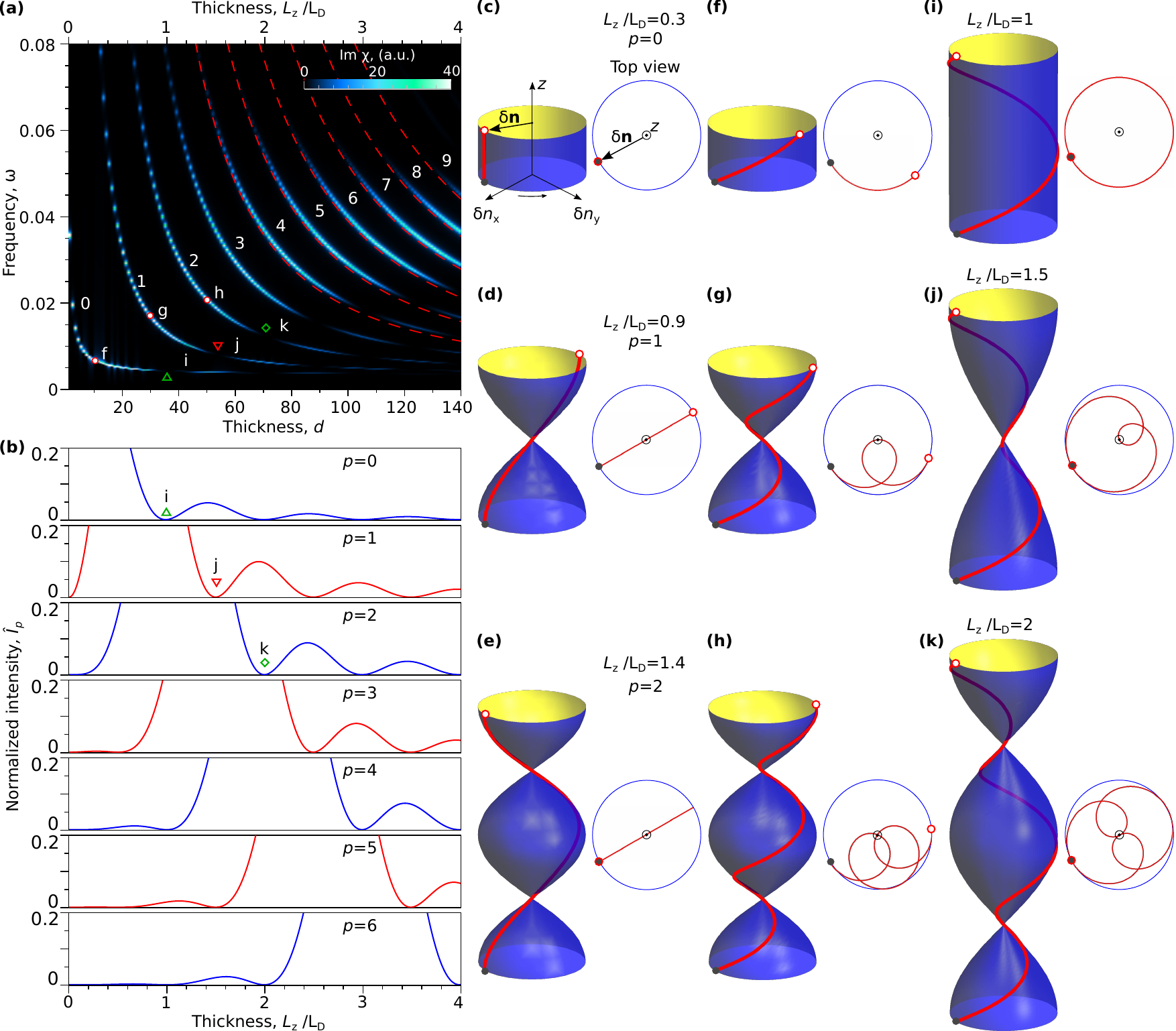}
\caption{(a) Absorption spectra $\mathrm{Im}\chi(\omega)$ for a film of chiral magnet in the field polarized phase ($\mathbf{B}_\mathrm{DC}\parallel \mathbf{e}_\mathrm{z}$, $B_{\mathrm{DC}}=1.1B_\mathrm{D}$) under in-plane excitation ($\mathbf{B}_\mathrm{AC}\perp \mathbf{e}_\mathrm{z}$).
The red dashed lines are the resonance frequencies \eqref{FMDFP} for modes with indexes $p=4$ - $9$.
The symbols indicate the corresponding figures with visualizations of dynamic magnetization profiles along the film thickness. 
(b) The normalised intensity of $\hat{I}_p=I_p/I_p^\mathrm{max}$ calculated with \eqref{Int} as a function of the film thickness for the modes with $p=0$ - $6$. 
(c)-(e) Dynamical magnetization profiles \eqref{FMSW} for the case of ferromagnetic film without DMI.
(f)-(h) Profiles of eigenmodes excited by in-plane uniform field $\mathbf{B}_\mathrm{AC}$ in magnetic film with DMI in field polarized phase. (i)-(k) Profiles of eigenmodes unexcited by in-plane uniform field $\mathbf{B}_\mathrm{AC}$ in magnetic film with DMI in field polarized phase.
The solid black circle and hollow red circle in (c)-(k) correspond the ends of the vectors $\langle\delta \mathbf {n}_{xy}\rangle$ at $z=0$ and $z=d$, respectively.
}
\label{FM_phase}
\end{figure*}

For the case of $\mathbf{B}_\mathrm{AC}\parallel \mathbf{e}_\mathrm{z}$, the profile of $\langle\delta \mathbf {n}(z)\rangle$ is also characterized by the presence of nodes and antinodes [Figs.~\ref{SKL}(i)-(k)] and thus can be thought of as SSW localized in the shells of skyrmion tubes.
The $\langle\delta \mathbf {n}(z)\rangle$ profile, in this case, does not have well-defined chirality as in the case of the CCW and CW modes excited by the in-plane field.
On the other hand, the $\langle\delta \mathbf {n}(z)\rangle$ profile is not flat as in the case of a pure ferromagnet.
The latter can be explained by the presence of the chiral surface twist~\cite{Rybakov_13, Rybakov_16} and the curvature of the skyrmion shell.

\section{Discussions}

Interesting aspect of the problem is the oscillatory fading and increasing of the dynamic susceptibility seen in the the low-frequency branches in the absorption spectrum for in-plane AC field [see Fig.~\ref{SKL}(a)].
To explain this phenomenon we use the standard approach for analysis of the spin waves~\cite{Melkov_book, Hoekstra_77, Salansky_book}.
Since the low-frequency modes of skyrmion lattice are localized in the regions where $\mathbf{n}_0\parallel\mathbf{e}_z$, we will consider the limiting case of the uniformly magnetized film and resonance of the spin-waves propagating in opposite directions along the $z$-axis.
Following the approach of Refs.~\cite{Garst_17, Zingsem_19} we derived the analytical equations for the dynamic magnetization of the eigenmodes: %
\begin{equation}
\begin{aligned}
    \delta n_x(z,t)= A\cos(Dz/J+\omega t)\cos(\pi p z/d),\\
    \delta n_y(z,t)= A\sin(Dz/J+\omega t)\cos(\pi p z/d),
\end{aligned}
\label{dmx_sw}
\end{equation}
where the term $\cos(\pi p z/d)$ defines the position of the nodes and antinodes, while $Dz/J$ is an additional phase shift across the film thickness (for details see Appendix \ref{Appendix_B}).
The resonance frequency of such standing spin-wave is a function of the thickness $d$ and the mode index $p$:
\begin{equation}
\omega_p=\left(\frac{p\pi}{d}\right)^2+\frac{\mu_\mathrm{s} (B_\mathrm{DC}-B_\mathrm{D})}{J}.\
\label{FMDFP}
\end{equation}

Figure~\ref{FM_phase}(a) shows a good agreement between the results obtained with spin-dynamics simulations and the analytical expression \eqref{FMDFP} for $B_\mathrm{DC}=1.1B_\mathrm{D}$.
The only exception is the limit of very thin films where the continuum model and discreet spin lattice model diverge~\cite{Rybakov_16}.

In Figs.~\ref{FM_phase}(c)-(k), we show a few representative examples of the dynamic magnetization profiles for different modes and thicknesses, which were calculated with \eqref{dmx_sw} and \eqref{FMDFP}.
The red lines represent the instant dynamic magnetization profiles at the fixed moment in time, and the surfaces are formed by a complete set of such snapshots over one period.
It is easy to see that analyticaly derived dynamical magnetization profiles in Figs.~\ref{FM_phase}(f)-(h) are consistent with that obtained in the numerical experiment for skyrmion lattice  under in-plane excitation [see Figs.~\ref{SKL}(c)-(e) and ~\ref{SKL}(f)-(h)].
For comparison, in Figs.~\ref{FM_phase}(c)-(e) we show the $\delta \mathbf {n}$ profiles corresponding to the case of SSW in pure ferromagnets, where at any fixed time, all the vectors of dynamic magnetization lay in the same plane.
From the projections of the $\delta \mathbf {n}$ profiles into the plane orthogonal to $z$-axis [Figs.~\ref{FM_phase}(c)-(e) on the right], it is seen that the total dynamic magnetization, $\int_{0}^{d} \delta n \,dz$, is nonzero only for $p=0$.
Because of that, in pure ferromagnets, the modes with $p>0$, strictly speaking, can not be excited by uniform AC field~\cite{Melkov_book}.
In chiral magnets, it is not the case because the $\delta \mathbf {n}$ profile is not flat.
On the other hand the surface of revolution created by the instant snapshots of the dynamical magnetization profiles are identical for chiral and flat standing spin-waves. Thereby, in the most general case, to identify the position of the nodes and antinodes one should take into account a complete 3D profile of the dynamical magnetization.
Othervise, when only one component of the dynamic magnetization is taking into account, one can come to a contradicting statement that the position of the antinodes is not fixed in time and thus the observed spin-waves can not be identified as standing~\cite{Zingsem_19}.
The variation of the projected on one plane dynamic magnetization of the chiral SSW is illustrated in Supplementary Movies 10.

To understand the reasons of the fading of the intensity of the chiral standing spin-waves one should take into account the total dynamic magnetization.
Note, the projection of the $\delta \mathbf {n}$ into the plane orthogonal to $z$-axis has a curved shape [Figs.~\ref{FM_phase}(f)-(h) on the right].
Because of that, the total dynamic magnetization is not zero, and chiral standing spin-waves can be easily excited even by the uniform AC field.
On the other hand, because of the additional phase shift in \eqref{dmx_sw}, the total dynamic magnetization is a function of the thickness.
In Figs.~\ref{FM_phase}(i)-(k) we provide three representative examples of the $\delta \mathbf {n}$ profiles at particular thicknesses at which projected $\delta \mathbf {n}$ represent closed loops and the total dynamic magnetization integrated over the thickness tends to zero.
The latter explains the fading zones in in the absorption spectrum, which are marked in Fig.~\ref{FM_phase}(a) by (i)-(k), respectively.
When the total dynamic magnetization of a particular mode tends to zero, the response of the system to the uniform AC field reduces, and we observe the fading of the intensity of that eigenmode in the absorption spectrum.

To quantify this effect we estimated the intensity $I_p$ of the resonance modes using the following well-known relation~\cite{Melkov_book, Hoekstra_77, Salansky_book}
\begin{equation}
I_p\sim \frac{\left(\int_{0}^{d} \delta n \,\mathrm{d}z\right)^2}{d \int_{0}^{d} \delta n^2 \,\mathrm{d}z},
\label{Int}
\end{equation}
where $\delta n\equiv\delta n(p,z,t,d)$  for the particular mode $p$ and thickness $d$ are defined by \eqref{dmx_sw}-\eqref{FMDFP}.
In Fig.~\ref{FM_phase}(b) we show the dependencies $I_p$ for the modes with index $p$ from 0 to 6. 
The position of the minima and maxima in $I_p(d)$ dependencies is fully consistent with the fading and increase of the $\mathrm{Im}\chi(\omega)$ in the simulated absorption spectrum in Fig.~\ref{FM_phase}(a).
One can conclude that the effect of the fading of the intensities of particular chiral SSWs originates from the chiral profile of the dynamic magnetization.
Noticeably, this effect also appears in the spectrum of the conical phase (see Appendix \ref{Appendix_C}).
The chiral profile of the dynamic magnetization in this case appear not because of the additional phase shift across the thickness but due to the chiral modulations in the static equilibrium state.

The above analytical approach can also be applied to the case of $\mathbf{n}_0\perp\mathbf{e}_z$, which can be compared to the modes excited in the skyrmion shell.
The dispersion relation $\omega(k)$ of spin-waves propagating along $z$-axis, in this case, is symmetric as in the pure ferromagnet~\cite{Cortes_2013}. 
In this case, the solutions corresponding to  SSWs are identical to those for pure ferromagnet and the dynamic magnetization profile is flat as in Figs.\ref{FM_phase}(c)-(e). 
Thereby,  SSWs are not chiral in this case.
For more details, see Appendix \ref{Appendix_B}.
On the other hand, in the skyrmion shell the static equilibrium magnetization across the film thickness is not ideally collinear and exhibits an additional chiral twist near the free surface of the sample~\cite{Rybakov_13, Rybakov_16}. 
Because of the modulation of $\mathbf{n}_0\equiv\mathbf{n}_0(z)$, similar to the case of the cone phase, the total dynamic magnetization becomes not zero, which allows exciting these modes with the uniform AC field.
Moreover, the magnetization in the skyrmion shell has a small out-of-plane component.
The spin-wave spectrum $\omega(k)$ in this case becomes asymmetric~\cite{Cortes_2013} and the dynamic magnetization $\langle\delta \mathbf {n}_{zy}\rangle$ get a weak twisting across the film thickness.
The twist induced by the magnetization with the positive and negative out-of-plane components has an opposite sign.
This effect is compensated in the ideal case of an isolated flat domain wall. 
However, since the shell of the skyrmion has a cylindrical shape, the volumes with the positive and negative out-of-plane magnetization are not identical, and that which corresponds to the outer region dominates.
The latter gives rise to a week twist of dynamical magnetization in the case of SSWs in the skyrmion shell.

\section{Conclusions}

In summary, we studied the eigenmodes of 3D skyrmion lattice in the film of the isotropic chiral magnet.
To illustrate the features of the spin-wave resonance in a 3D skyrmion lattice, we calculated the absorption spectra for the cases of in-plane and out-of-plane excitations.
In the case of in-plane excitation, the absorption spectrum calculated as a function of the film thickness is characterized by a set of low-frequency and high-frequency modes appearing in pairs. 
In this case, the maxima of the spin-wave excitations are localized in the inter-skyrmion area or in the skyrmion core, where magnetization is normal to the film plane.
We identify the corresponding resonance modes as chiral standing spin waves.
In contrast to standard standing spin waves in ferromagnets without DMI, the profiles of dynamic magnetization of chiral standing spin waves are characterized by helical modulation with the pitch determined by the competition between DMI and Heisenberg exchange.
An important consequence of such modulation of dynamic magnetization is the periodic fading of the absorption spectra intensity with the variation of the plate thickness.
A detailed analysis revealed that this effect takes place not only in the 3D skyrmion lattice but also in the case of the filed saturated state and the conical phase excited by the uniform field. 

Under out-of-plane excitation, the absorption spectrum also demonstrates the appearance of standing spin waves, which are localized in the skyrmion shell where spins lie in the plane of the plate. 
The helical modulations of the dynamic magnetization profile are not present. Thus, standing spin waves, in this case, have more in common with standard standing spin waves in pure ferromagnets.
We show that the effect of the chiral surface twist plays an essential role in this case.

\section*{Supplementary Material}
See supplementary Movies 1-9 [URL:] for more information about rotation of dynamic magnetization vectors from Figs.~\ref{SKL}(c)-(k). The supplementary Movie 10 [URL:] demonstrates the variation of dynamic magnetization in 3D space compared to its 2D projection for the case of Fig.~\ref{FM_phase}(g). 

\begin{acknowledgments}
Authors thank S.V. Tarasenko for fruitful discussion. 
The authors acknowledge financial support from the European Research Council (ERC) under the European Union’s
Horizon 2020 research and innovation program (Grant No. 856538, project “3D MAGiC”).
F.N.R. acknowledges support from the Swedish Research Council.
\end{acknowledgments}

\appendix

\section{Methods}\label{Appendix_A}

\subsection{Parameters of the model  and spectra calculation}

The LLG equation (\ref{LLG1}) for underlying Hamiltonian (\ref{E_tot}) was numerically solved by the fourth-order Runge-Kutta method.
The spectra were calculated with Excalibur code~\cite{Excalibur} and double-checked with the publicly available code Mumax~\cite{Mumax}.
The modes visualization were performed with the code Magnoom~\cite{magnoom}.

The following parameters were used: $B_\mathrm{DC}=0.5B_\mathrm{D}$, $D/J=0.18$ ($L_\mathrm{D}=34.91a$), $\alpha=0.01$ and the time step  $dt=0.05$.
To excite the oscillations in the given ground state, we follow the standard approach~\cite{Marty_02, Mochizuki_2012}.
First, we excite the system with the time-step function~\cite{Marty_02} with the amplitude $B_\mathrm{AC}=0.03B_\mathrm{DC}$.
After the excitation field is switched off at $t=0$, we record the total magnetization of the system at each iteration.
To ensure high resolution of the spectra, the total number of iteration is set to~${2^{19}\approx 5\times 10^5}$.
Using the discrete Fourier transform of $\mathbf{n}(t)$, we  calculate the dynamic magnetic susceptibility  $\chi(\omega)$ as a function of oscillation frequency $\omega$. 
In the absorption spectra presented in the main text we show the imaginary part of dynamic susceptibility $\mathrm{Im}\chi(\omega)$, for different film thicknesses.
We also tested the magnetization excitation by sinc function~\cite{Beg_17, Fangohr}, and found that, in this case, it does not qualitatively influence the $\mathrm{Im}\chi(\omega)$ dependencies.
As follows from (\ref{LLG1}) the unitless frequency $\omega$ is related to frequency $f$ in Hz as  $\omega=f(2\pi\mu_s)/(\gamma J)$. For instance, for $J=0.4$ meV, $\omega=0.01$  corresponds  $f \approx 1$GHz. 

\subsection{Calculation of the distribution and profile of the dynamic magnetization}
To identify the distribution of dynamic magnetization $\delta \mathbf n$ depicted on the left panel of  Figs.~\ref{SKL}(c)-(k) and corresponding profiles of averaged dynamic magnetization $\langle\delta \mathbf {n}\rangle(z)$, the relaxed 3D SkL was excited by AC magnetic field $B_\mathrm{AC}=A\sin \omega_p t $ at corresponding resonance frequency $\omega_p$.
Since we consider small oscillations, the amplitude of exciting field $A = 0.01B_\mathrm{DC}$ for $\mathbf{B}_\mathrm{AC}\parallel \mathbf{e}_\mathrm{x}$ and $A = 0.00125B_\mathrm{DC}$ for $\mathbf{B}_\mathrm{AC}\parallel \mathbf{e}_\mathrm{z}$. 
To reach the dynamic equilibrium regime, we perform simulation in the long time range.
After that, we took the snapshots of the entire magnetization vector field with the time interval $\Delta t= 2\pi/(\omega_p N)$, where $N=32$ is a number of the snapshots.
To ensure precision in these calculations, we fix the time step $dt$ in the LLG simulation to be multiple to $\Delta t$, meaning $\Delta t/dt$ is always an integer.
%
%
In addition, the snapshots of the magnetization are averaged over ten periods.
Then, for each snapshot, we calculate the normalized dynamic component of magnetization at each $i$-th site, $\delta \mathbf n_i(t)=(\mathbf n_i(t)-\mathbf{n}_0)/\delta n_\mathrm{max}$, where $\delta n_\mathrm{max}=\mathrm{max}|\mathbf n_i(t)-\mathbf{n}_0|$ is the maximal dynamic magnetization over the whole sites and whole period of oscillation, \textit{i.e.} $|\delta \mathbf n_i| \leqslant1$ for any $t$ and $i$.
%
The normalized that way vector field $\delta \mathbf n_i(t)$ can be visualized using the color scheme explained in  Fig.~\ref{Fig1}(c).
In the left panel of Figs.~\ref{SKL}(c)-(k), we show representative example of the $\delta \mathbf n$ distribution over the simulated domain at fixed moment in time. 

\begin{figure*}
\centering
\includegraphics[width=17.0cm]{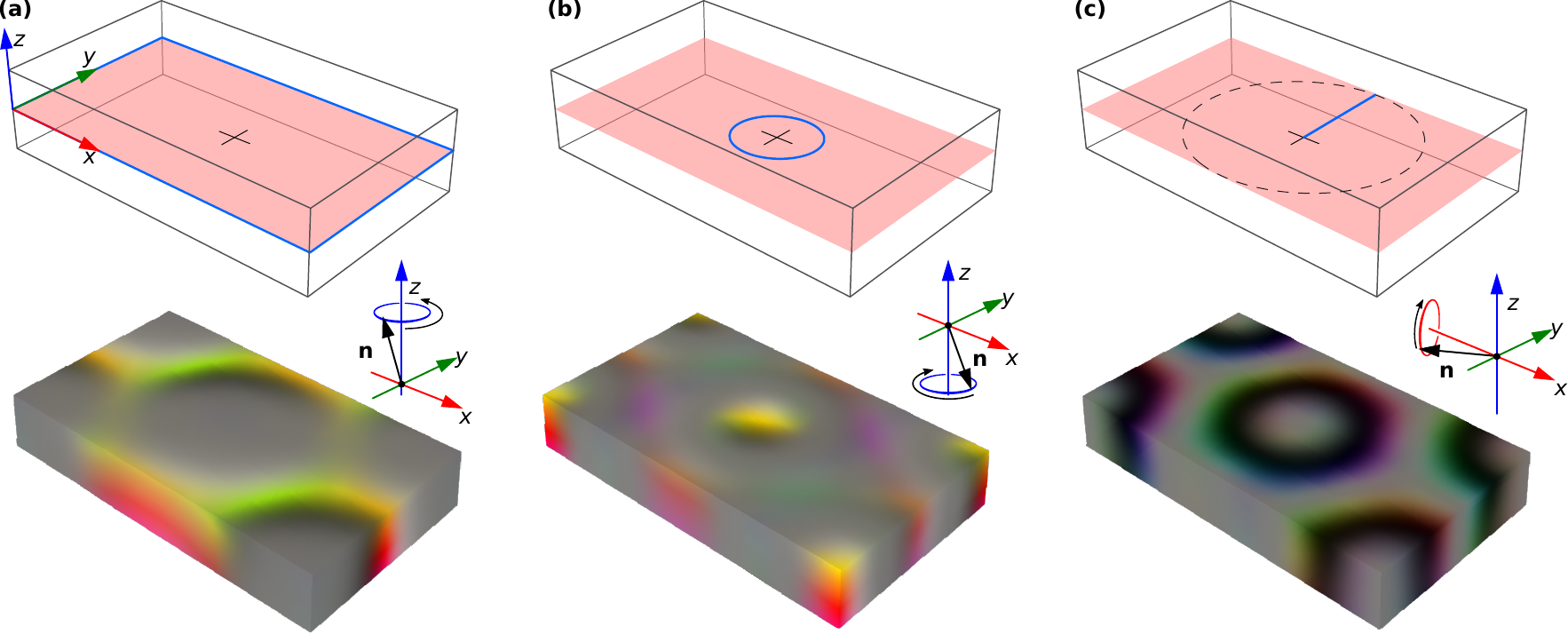}
\caption{The area $\Omega$ in the $xy$-plane for the averaged dynamic magnetization, $\langle\delta \mathbf{n}\rangle(t,z)=\sum_{(x,y)\in\Omega}\delta \mathbf{n}(t,\mathbf{r})$, for different modes.
(a) and (b) correspond to the case of $\mathbf{B}_\mathrm{AC}\perp \mathbf{B}_\mathrm{DC}$ for low-frequency and high-frequency modes, respectively. (c) corresponds to the case of $\mathbf{B}_\mathrm{AC}\parallel \mathbf{B}_\mathrm{DC}$.
The top row of images show the simulated domain, the arbitrary chosen $xy$-plane and the area $\Omega$ bounded by the blue line.
The bottom row of images shows the snapshots of the $\delta \mathbf{n}$ distribution provided for the reference, see also Figs.~\ref{SKL}(c), (f), and (i).
In (a) $\Omega$ corresponds to the whole $xy$-plane, in (b) $\Omega$ is bounded by the circular domain of the radius corresponding to the lowest value $|\delta \mathbf{n}|$ around the skyrmion core, and in (c) $\Omega$ corresponds to the line segment drawn from the center of the simulated domain and parallel to the $y$-axis. 
}
\label{FigS-modes}
\end{figure*}

Because the magnetization precession axis and localization of the modes are different [Fig.~\ref{SKL}], the area $\Omega$ for calculation of the average dynamic magnetization profiles have been chosen differently, as illustrated in Fig.~\ref{FigS-modes}. 
For the low-frequency modes, Figs.~\ref{SKL}(c)-(e), this area occupies the whole $xy$-plane in the simulation domain, $\Omega=L_\mathrm{x}\times L_\mathrm{y}$.
Although $\delta\mathbf{n}$, in this case, is mainly localized in the inter-skyrmion region, the contribution from other parts of the domain is negligibly small, and for simplicity, the averaging area $\Omega$ can be extended to the whole domain. 

For the high-frequency modes, Figs.~\ref{SKL}(f)-(h), we bound the averaging area by a disk, $\Omega=2\pi R^2$, where $R$ is defined as a distance from the skyrmion center at which the dynamic magnetization reaches minima, $\delta \mathbf{n}(|\mathbf{r}|=R)\rightarrow\mathrm{min}$.
Such a choice of $\Omega$ allows us to exclude small excitations that appeared in the inter-skyrmion region because of the coupling to more intense low-frequency modes. 

In the case of the in-plane AC field, the dynamics magnetization mainly lies in the $xy$-plane ($\delta n_\mathrm{z}$ component is negligibly small).
We assume that the normal to the $xy$-plane is parallel to  $\mathbf{e}_\mathrm{z}$ in both cases of low-frequency and high-frequency modes.
As seen from Figs.\ref{FigS-modes}(a) and \ref{FigS-modes}(b), 
in this case, the projected into $xy$-plane dynamics magnetization precesses CCW and CW, respectively. 

The contrary to above, for modes excited by the out-of-plane AC field [Figs.~\ref{SKL}(i)-(k)], there is no common precession axis for the spins lying in the skyrmion shell. 
In this case, we choose the area $\Omega$ as a line segment connecting the skyrmion center and an arbitrary point on the circle of radius $L_\mathrm{y}/2$ around it.
A particular choice of the point on the circle does not play a role.
For definiteness, we select that line segment such that it is parallel to the $y$-axis, as shown in Fig.~\ref{FigS-modes}(c).
In this case, the dynamics magnetization projected to the $yz$-plane  exhibits CCW rotation because we select the normal to the $yz$-plane to be along $-\mathbf{e}_\mathrm{x}$. Otherwise, the profiles of the average dynamics magnetization would precess CW.

\begin{figure*}
\centering
\includegraphics[width=15.0cm]{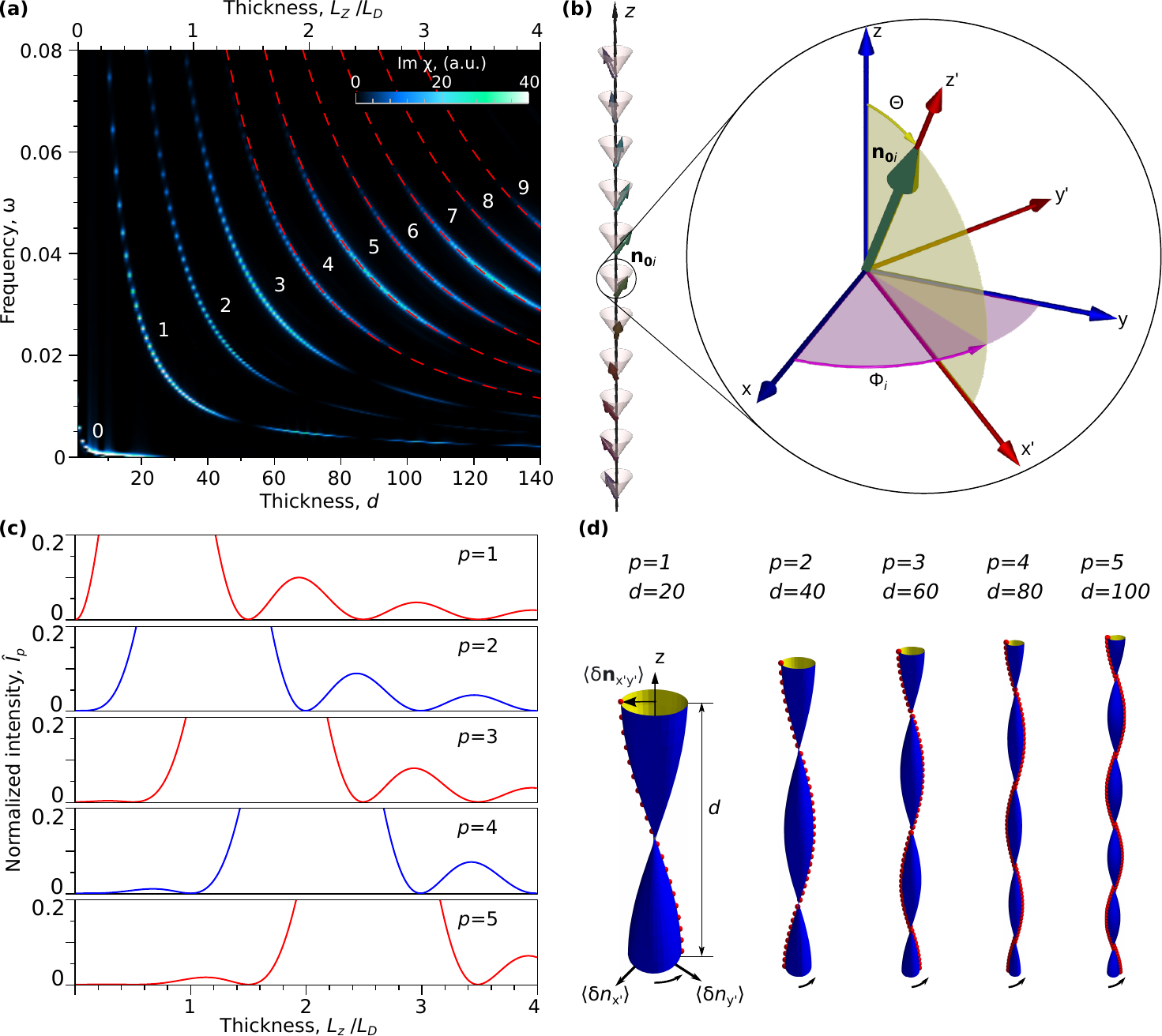}
\caption{(a) Absorption spectra $\mathrm{Im}\chi_{xx}(\omega)$ of chiral magnetic film in the conical spin spiral state for different thicknesses $d=L_z/a$, (dark-blue-white color map) and analytical spectrum \eqref{Sp_SSW} (red dashed lines) for $B_{\mathrm{DC}}=0.85B_\mathrm{D}$; white digits correspond to indexes of modes. (b) The equilibrium magnetic ordering $\mathbf{n}_0$ in conical spin spiral state along film thickness and the local coordinate system $x'y'z'$ associated with $\mathbf{n}_{0 \mathit{i}}$ in the each $i$-th layer ($z' \parallel \mathbf{n}_{0 \mathit{i}}$). (c) The values of \eqref{Int} as a function of the film thickness for modes with indexes $p=1,2,3,4,5$ in the conical spin spiral state. (d) Snapshots of dynamic magnetization $\langle\delta \mathbf {n}_{x'y'}\rangle$ in local system $x'y'z'$ at fixed time moment for modes with indexes $p=1,2,3,4,5$.  
}
\label{cone_SSW}
\end{figure*}
\section{Resonance eigenmodes of saturated state}\label{Appendix_B}

%
Here we recall the results of Refs.~\cite{Garst_17, Zingsem_19} and show that the analytical solutions for the profile of dynamic magnetization are fully consistent with the results of our spin-dynamic simulations.
We consider the case when the magnetization $\mathbf{n}_0$ in static equilibrium is parallel to $\mathbf{e}_\mathrm{z}$ in a whole sample. 
The spin wave propagates along $z$ axis.
Since the amplitudes of dynamic magnetization $\delta \mathbf {n}$ are  small, one can take into account only orthogonal components $\delta n_x$ and $\delta n_y$ of dynamic magnetization and neglect the $\delta n_z$ component.
For unconfined crystal this leads to the following dispersion relation~\cite{Garst_17}:
\begin{equation}
\omega=  \left(\overline{k}^2 \pm \dfrac{2D}{J} |\overline{k}|  +\dfrac{\mu_\mathrm{s} H}{J}\right),
\label{FMD}
\end{equation}
where $\overline{k}=ka$ and $\omega$ are the absolute values of the dimensionless wave vector and the frequency of a spin-wave, respectively.
$H$ in \eqref{FMD} is absolute value of the effective (internal) magnetic field which in static equilibrium $\mathbf{H}_{i} \parallel \mathbf{n}_{0i}$ at any site. The sign "$\pm$" in \eqref{FMD} defines the spin-wave propagation direction.
In the limiting case of $D=0$, the dispersion relation \eqref{FMD} converges to the  well-known solution for a bulk ferromagnet,
\begin{equation}
\omega= \overline{k}_\mathrm{FM}^2+\mu_\mathrm{s} H/J,
\label{FMND}
\end{equation}
where $\overline{k}_\mathrm{FM}$ is the wave vector in pure ferromagnet.
Using circular variable, $ \delta n = \delta n_x + i\delta n_y$, the superposition of two spin waves propagating in the opposite directions along the $z$-axis in the case of chiral magnet can be written as 
\begin{equation}
\delta n(z,t)= \frac12A\exp(i\omega t)(\exp(i\overline{k}_r z)+\exp(i\overline{k}_l z)),
\label{DMSW}
\end{equation}
where $\overline{k}_r$ and $\overline{k}_l$ are the wave vectors for spin waves propagating parallel and anti-parallel $z$-axis.
In the case of a pure ferromagnet the dynamic magnetization corresponding to the superposition of two waves takes the form
\begin{equation}
\delta n(z,t)= \frac12A\exp(i\omega t)(\exp(i\overline{k}_\mathrm{FM} z)+\exp(-i\overline{k}_\mathrm{FM} z)).
\label{FMSW}
\end{equation}
Using \eqref{FMD}-\eqref{FMSW}, the vectors $\overline{k}_r$ and $\overline{k}_l$  can be written as $\overline{k}_r=\overline{k}'+\overline{k}_\mathrm{FM}$ and $\overline{k}_l=\overline{k}'-\overline{k}_\mathrm{FM}$. 

As follows from \eqref{FMD}, for small $D/J$, the wave vector shift, up to the first leading term, is $\overline{k}'= D/J$. 
%
Let us assume that the sample has a shape of a film in $xy$-plane with the boundaries located at $z=0$ and $z=d$.
The boundary conditions are~\cite{Garst_17} 
\begin{equation}
\partial_\mathrm{z}\mathbf{n}-(2 \pi/L_\mathrm{D}) \textbf{e}_\mathrm{z}\times \mathbf{n}|_{z=0,d}\,=0. 
\label{Bnd_con}
\end{equation}
%
%
By substituting \eqref{DMSW} in \eqref{Bnd_con} we obtain the set of solutions with nonzero amplitudes, $\overline{k}_\mathrm{FM}=\pi p/d$, where $p=0,1,2,...$.
Then, the in-plane components of the dynamic magnetization for the SSW takes the form of \eqref{dmx_sw} in the main text.

Now let us consider the case when the magnetization lies in the film plane, $\mathbf{n}_0 \perp\mathbf{e}_\mathrm{z}$.
For definiteness we will assume that $\mathbf{n}_0 \parallel \mathbf{e}_\mathrm{x}$ [see Fig.~\ref{FigS-modes}(c)].
For spin waves propagating along the $z$-axis, the dispersion relation $\omega(k)$ is symmetric~\cite{Cortes_2013}.
In the case of unconfined crystal of chiral magnet the the dispersion relation has the form \eqref{FMND}. 
For the plate, using \eqref{FMSW} and boundary conditions \eqref{Bnd_con}, we get the solution for the SSW with the ``flat'' profile of $\langle\delta \mathbf {n}_{zy}\rangle$
\begin{equation}
\begin{aligned}
    \delta n_y(z,t)= A\sin(\omega t)\cos(\pi p z/d),\\
    \delta n_z(z,t)= A\cos(\omega t)\cos(\pi p z/d),
\end{aligned}
\label{dm_OOFP}
\end{equation}
Thereby, the SSW is not chiral in this case. The same is true for SSW localized in the shell of skyrmion.

\section{Absorption spectra and resonance eigenmodes of conical spin spiral state}\label{Appendix_C}
Let us consider the case when $\mathbf{n}_0$ in static equilibrium represents a cone phase with the wave vector $Q=2 \pi/L_{\mathrm{D}}$ parallel to $z$-axis,  $\mathbf{n}_0= \left( \sin\Theta \cos\Phi, \sin\Theta \sin\Phi, \cos\Theta    \right)$,
where $\Theta$ and  $\Phi=Qz$ are polar and azimuthal angles, respectively. 
For definiteness, in the following we assume that the static magnetic field $\mathbf{B}_\mathrm{DC}\parallel \mathbf{e}_\mathrm{z}$ and  $B_{\mathrm{DC}}=0.85B_\mathrm{D}$ and the conical phase is the ground state of the system in a wide range of film thickness\cite{Rybakov_16}.
When spin wave propagate along $\mathbf{e}_\mathrm{z}$, the dispersion relations for bulk chiral magnet has the form \cite{Garst_17}
\begin{eqnarray}
{\omega}=|\overline{k}| \sqrt{\overline{k}^2+\overline{Q}^2 \cdot \sin^2(\Theta)},
\label{Sp_prop_con}
\end{eqnarray}
where $\overline{Q}=Qa$. 
Using \eqref{Bnd_con}, one can show that in the eigenfrequency of the SSW is~\cite{Garst_17}
\begin{eqnarray}
{\omega}_p(d)=\dfrac{p\pi}{d} \sqrt{\left(\dfrac{p\pi}{d}\right)^{2}+\left(\dfrac{D}{J}\right)^2  \left(1-\dfrac{B_\mathrm{DC}^2}{B_\mathrm{D}^2}\right)},
\label{Sp_SSW}
\end{eqnarray}
which remains valid only for ${B_\mathrm{DC} \leqslant B_\mathrm{D}}$.
Figure~\ref{cone_SSW}(a) shows the absorption spectrum $\mathrm{Im}\chi_{xx}$ for cone phase obtained in numerical simulations as a function of the film thickness.
The eigenfrequencies of the SSWs according to Eq.~\eqref{Sp_SSW} show good agreement with spin-dynamics simulations.
Similar to the case of the spectrum for the field saturated state in Fig.~\ref{FM_phase}, the only exception is the limit of very thin films where the continuum model and discreet spin lattice model diverge~\cite{Rybakov_16}.
According to Eq.~\eqref{Sp_SSW}, the eigenfrequency $\omega_0=0$ at any thickness, while in numerical simulations at small thicknesses it has a nonzero value. 

Similar to the case of saturated state the absorption spectra for the cone phase are also characterized by periodical fading of the mode intensities.
To make the periodical decay of $\mathrm{Im}\chi_{xx}$ more visible, we show the analytical dependencies (dashed red lines) only for $p\geq4$.
In contrast to the saturated state \eqref{FMD}, the $\omega(k)$ relation for the cone phase \eqref{Sp_prop_con} is symmetric.
The SSW in this case are formed by two opposite propagating spin waves with identical $k$-vector.
%
%
To explain the presence of the fading zones following with the period of cone modulations, $L_\mathrm{D}$ in Fig.\ref{cone_SSW}(a), one should take into account the spiraling of the magnetization in static equilibrium state [Fig.\ref{cone_SSW}(b)].
To take this into account, we apply the rotation matrix associated with cone phase to the two counterpropagating spin-waves in ferromagnetic state \eqref{FMSW}.
Then, by substituting the obtained $\delta \mathbf{n}(z,t)$ into \eqref{Int}, we calculate the dependencies of the eigenmode intensities as functions of the film thickness, which are depicted in Fig.~\ref{cone_SSW}(c). The position of the minima and maxima in Figs.~\ref{cone_SSW}(a) and ~\ref{cone_SSW}(c) show excellent agreement.
Interestingly, in the coordinate system related to the static equilibrium state of the cone phase illustrated in Fig.~\ref{cone_SSW}(b) the $\delta \mathbf{n}$ profiles of SSW are flat [Fig.~\ref{cone_SSW}(d)], as in the case of eigenmodes of pure ferromagnet.
The profiles of $\delta \mathbf{n}$ in Fig.~\ref{cone_SSW}(d) are obtained in numerical simulations.
The coordinate transformation depicted in Fig.~\ref{cone_SSW}(b) effectively unwind the cone phase into the state with $\mathbf{n}_0\parallel\mathbf{e}_\mathrm{z}$. Thus the non zero intensities of the SSW exited by the uniform AC field in our numerical experiment are solely explained by the twist of magnetization $\mathbf{n}_0$ in the static equilibrium state.
The fading of the intensities following with the period of cone modulations, $L_\mathrm{D}$, appears at the thickness which satisfy the criteria $I_p(d)\rightarrow0$ in \eqref{Int}.


\end{document}